\documentstyle[pre,aps,epsfig,multicol]{revtex}
\begin{document}
\draft
\widetext

\newcommand{\be}{\begin{equation}}
\newcommand{\ee}{\end{equation}}
\newcommand{\ber}{\begin{eqnarray}}
\newcommand{\eer}{\end{eqnarray}}

\title{Empirical Phase Diagram of Congested Traffic Flow}

\author{H. Y. Lee$^{1,2}$, H.-W. Lee$^{1}$, and D. Kim$^{1,2}$}
\address{
$^{1}$Center for Theoretical Physics, Seoul National University, 
Seoul 151-742, Korea \\
$^{2}$Department of Physics,
Seoul National University, 
Seoul 151-742, Korea }

\maketitle

\begin{abstract}
We present an empirical phase diagram of the congested 
traffic flow measured on a highway section with one effective
on-ramp.
Through the analysis of local density-flow relations
and global spatial structure of the congested region, 
four distinct congested traffic states are identified.
These states appear at different levels of the upstream
flux and the on-ramp flux, thereby generating a phase digram of the
congested traffic flow.
Observed traffic states are discussed in connection with recent 
theoretical analyses.
\end{abstract}
\widetext

\begin{multicols}{2}

\narrowtext
%\widetext

 Rich physical phenomena in traffic flow have been
investigated by both empirical and theoretical studies.
Two distinct traffic states, the free flow and 
the traffic jam state,
have been identified from measurements on homogeneous highways
\cite{Kerner96E} and
various properties of them
have been successfully reproduced by traffic models 
\cite{KNagel}.

Further empirical studies \cite{Kerner96E1,Kerner97L}
have reported the presence of additional traffic states, where
typical vehicle velocities take intermediate values
between the velocity levels of 
the free flow and the traffic jam.
In these states, vehicle motions in
all lanes are synchronized and
the notion of a unique density-flow relation breaks down.
From local traffic patterns, three different types
are classified \cite{Kerner96E1}. 
These congested traffic states \cite{CTcomment} appear mostly near ramps 
and theoretical studies \cite{Lee98L,HelbingL} of the ramp effects 
reproduced some of their features 
such as the discontinuous transition from the free flow to the congested flow
and the high flux level of the congested flow. 

 Recently, more extensive theoretical studies \cite{Helbingcondmat,Lee99E}
have revealed that for highways with an on-ramp,
several distinct congested traffic states appear
depending on the levels of the upstream flux and the on-ramp flux.
A phase diagram is constructed \cite{Helbingcondmat,Lee99E} and
metastabilities between these states are also investigated \cite{Lee99E}.
Yet the relation between these studies and the observations
\cite{Kerner96E1,Kerner97L} is not clear and especially
no empirical evidence for the predicted phase diagram has been
reported to our knowledge. 

 In this Letter, we present
an analysis of the congested traffic flow
measured on a highway in Seoul, Korea. 
From the studies of the traffic patterns at fixed locations
and the global spatial structures, we find 4 qualitatively distinct 
congested traffic states. 
Appearance of each state
depends on the upstream flux and the on-ramp flux,
from which we construct the empirical phase diagram. 
These 4 states exhibit features that agree 
with theoretical studies \cite{Helbingcondmat,Lee99E}.
But differences are also discovered.

\begin{figure}
\begin{center}
\epsfig{file=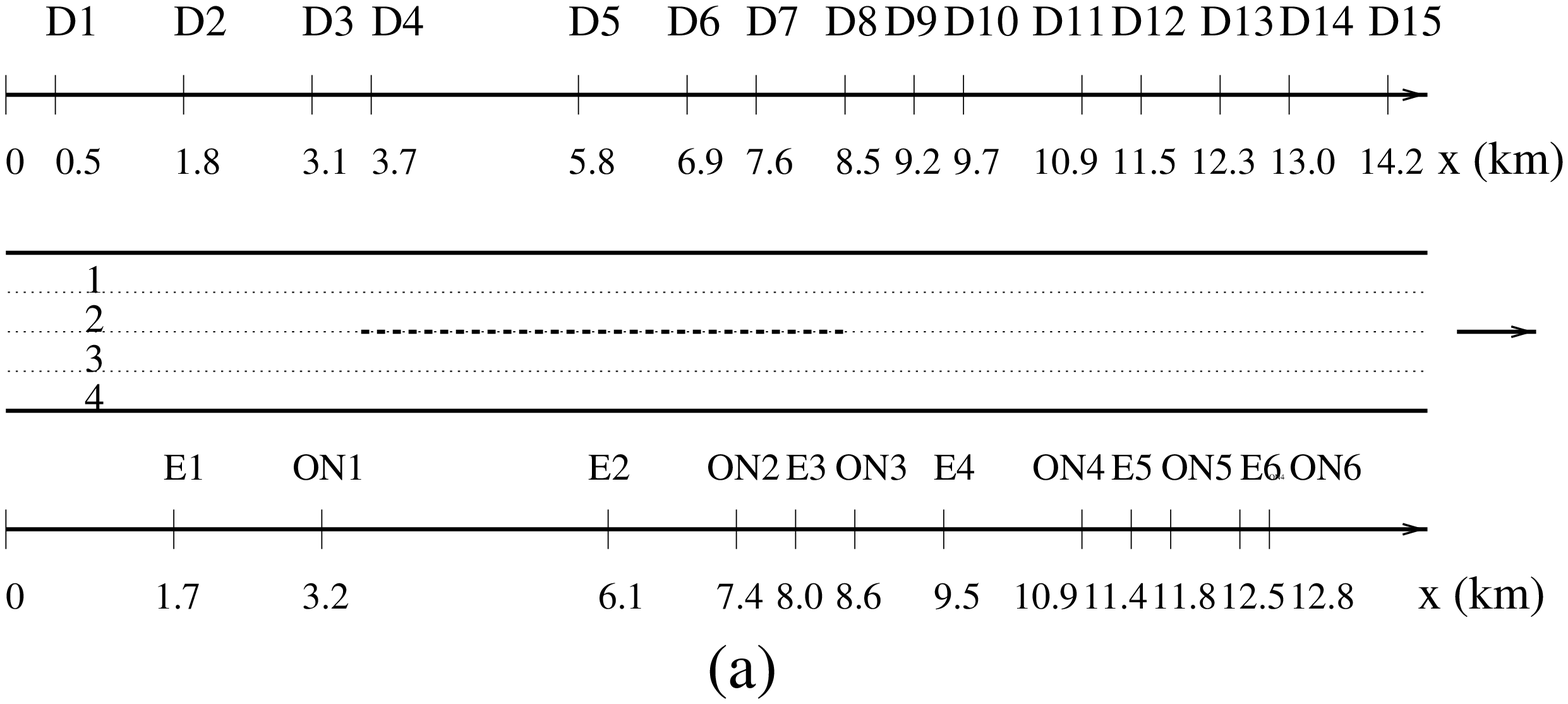,clip=,width=0.95\columnwidth}
\epsfig{file=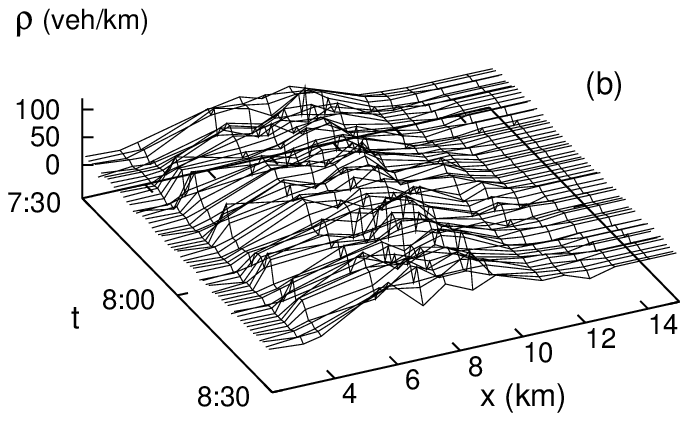,clip=,width=0.95\columnwidth}
\epsfig{file=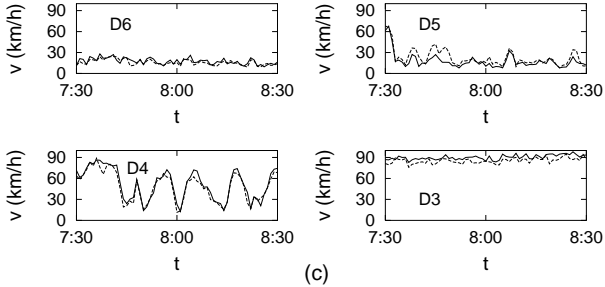,clip=,width=0.95\columnwidth}
\epsfig{file=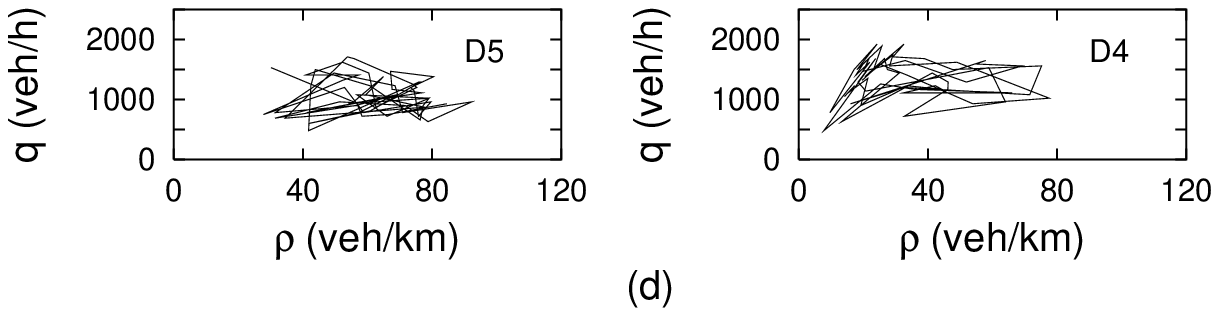,clip=,width=0.95\columnwidth}
\end{center}
\caption[]{
(a) Detector (D$n$), on-ramp (ON$n$), and off-ramp (E$n$)
locations in the Olympic Highway section studied in this work.
The dashed line in the middle denotes the lane divider.
(b) The 3D density profile of the CT1 state.
(c) Velocity vs. time plots at different detectors
(solid line for the lane 1 and dashed line for the lane 2).
The stop-and-go pattern appears at D4 but the free flow is observed at D3.
(d) The density-flow relations at D5 and D4 from 7:30 to 8:30 data.
The characteristic line of the traffic jam does not appear.
}
\label{fig:roadconf}
\end{figure}

 For the study, we use the traffic data of the Olympic Highway 
which connects the east and west ends of Seoul.
Since it is an intra-city highway, there are many ramps
and the spacings between them
are rather short ($\sim$ 1 km). But in some sections,
lane dividers separate the lane 1 and 2 from the lane 3 and 4,
and as a result, the two inner lanes become a ramp-free highway.
Our investigation is focused on a 14 km east-bound section,
between Seoul bridge and Young-dong bridge, that contains
a long ($\approx$ 5 km) lane divider \cite{dividercomment}.
In this section, 6 on-ramps and 6 off-ramps exist
(all of them connected to the outermost lane 4) and 15 image
detectors record the flux $q$ and the average velocity $v$
for each lane at every one minute interval [Fig.~\ref{fig:roadconf}(a)].
Among the ramps located outside
the ideal ramp-free region, usually only one or two of them 
are effective and the flux through other ramps are small.
For the analysis, inner-lane traffic data for 14 different days 
in June and July, 1998 are used.
We search congested traffic states that are long-lived ($\sim$ 1 hour) 
and appear practically everyday.
In this way, we identify 4 congested traffic states, 
for each of which, data for a particular day are presented
below to demonstrate its typical features.

 The first kind of the congested traffic states, which we call CT1, is shown 
in Fig.~\ref{fig:roadconf}(b), where the on-ramp ON3 at $x=8.6$ km 
is the main ramp in the time interval depicted in the figure
and the traffic between $x\approx 3$ km and $x\approx 10$ km 
is congested.
The spatial extension of the congested region does not expand
with time (within our estimated accuracy of 1 km/h) but
the boundaries of the region are not stationary.  
Also systematic oscillation develops spontaneously
near the upstream boundary, 
which is manifest in the velocity vs. 
time plot at D4 [Fig.~\ref{fig:roadconf}(c)].
Notice that the velocity oscillation is amplified and 
its period is enhanced 
as the upstream boundary is approached.
Peaks of the velocity move with velocity $\sim -$13 km/h.
Similar features are reported in Ref.~\cite{Kerner98L}.
We mention that the oscillation is not caused
by the temporal variations in the upstream flux.
Another charateristic of the CT1 state is the two-dimensional
covering of the density-flow relation [Fig.~\ref{fig:roadconf}(d)].

\begin{figure}
\begin{center}
\epsfig{file=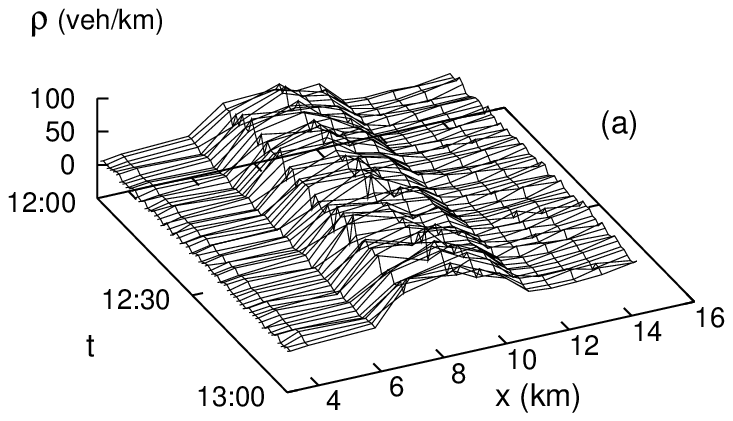,clip=,width=0.95\columnwidth}
\epsfig{file=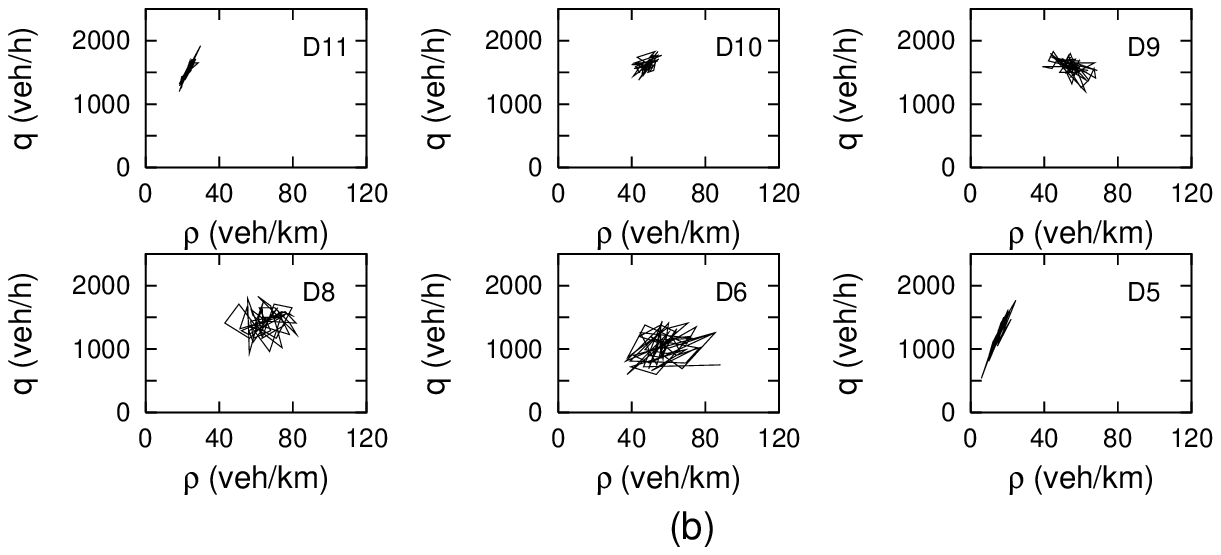,clip=,width=0.95\columnwidth}
\epsfig{file=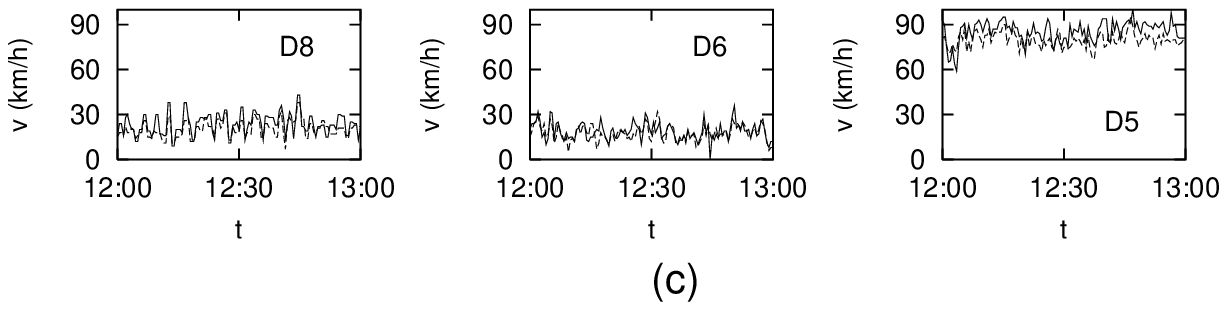,clip=,width=0.95\columnwidth}
\end{center}
\caption[]{
The 3D density profile (a), density-flow relations (b), and
velocity vs. time plots (c) of the CT2 state.
In contrast to the CT1 state, the amplification of the velocity 
fluctuations does not occur.
}
\label{fig:CT2}
\end{figure}

 The second kind of the congested traffic states, CT2, is shown 
in Fig.~\ref{fig:CT2}(a), where the effective ramp is ON3. 
The density-flow relations cover two-dimensional areas [Fig.~\ref{fig:CT2}(b)].
In contrast to the CT1 state, however, the boundaries of the congested 
region are almost motionless  and 
the development of the large amplitude oscillation
is not observed  [Fig.~\ref{fig:CT2}(c)] even though
the velocity levels are comparable to those
in the congested region of the CT1 state.

\begin{figure}
\begin{center}
\epsfig{file=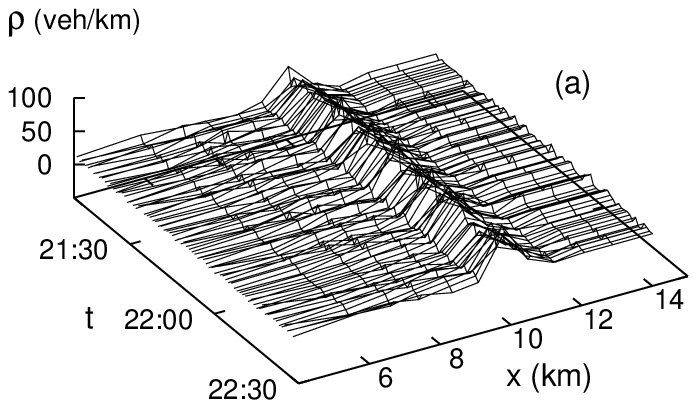,clip=,width=0.95\columnwidth}
\epsfig{file=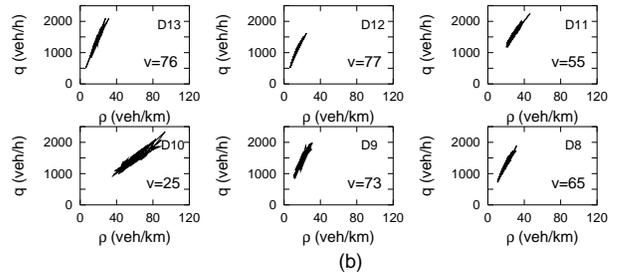,clip=,width=0.95\columnwidth}
\end{center}
\caption[]{
The 3D density profile (a) and density-flow relations (b) of
the CT3 state.
The velocity remains almost constant at each detector.
}
\label{fig:CT3}
\end{figure}

 The third kind of the congested traffic states, CT3, is shown in 
Fig.~\ref{fig:CT3}(a), where the effective ramp is ON4.
The boundaries of the congested region are again motionless.
The congested region is much shorter compared to the CT1 and CT2
states.
More important difference appears in Fig.~\ref{fig:CT3}(b).
The density-flow relation at each detector location inside the congested region
forms a straight line, implying that
the velocity ($v=q/\rho$) remains almost constant
even under the significant flux fluctuations.
However the values of the velocity are different for
different detectors. 
[In 3 out of 14 investigated days, we also observe
a wide congested region ($\sim$ 4 km) with the stationary and almost
{\it homogeneous} velocity profile.
Here we do not present this as another distinct traffic state
since the number of available data sets is too small.]

\begin{figure}
\begin{center}
\epsfig{file=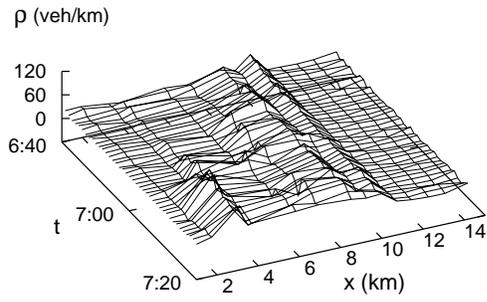,clip=,width=0.95\columnwidth}
\end{center}
\caption[]{
The 3D density profile of the CT4 state.
}
\label{fig:HCT}
\end{figure}

 The fourth and the last kind of the congested traffic states, CT4, is shown in
Fig.~\ref{fig:HCT}, which appears during
morning rush hours. The effective ramp is ON3.
While the downstream boundary of the congested region remains stationary,
the upstream boundary propagates backwards and the congested region 
expands monotonically, unlike all other congested traffic states mentioned 
above.
The expansion rate is higher for the higher level of the 
flux in the upstream region where the free flow is maintained, and
the observed values of the rate range from 2.2 km/h to 8.8 km/h.
The density-flow relation covers a two-dimensional area 
and the development of the large amplitude
oscillation
is not observed.

\begin{figure}
\begin{center}
\epsfig{file=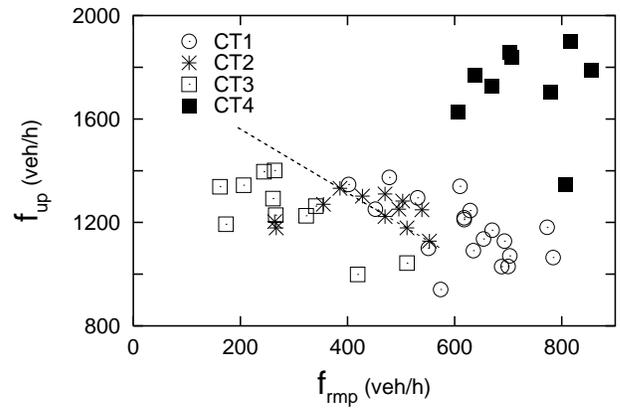,clip=,width=0.95\columnwidth}
\end{center}
\caption[]{
The phase diagram of the four congested traffic states on the Olympic Highway.
The upstream flux $f_{\rm up}$ and the on-ramp flux $f_{\rm rmp}$
are average values over the time interval during which a particular 
congested traffic state is maintained.
The free flow is also observed below the dashed line.
}
\label{fig:phase diagram}
\end{figure}

  We examine differences in the appearance conditions of the 4 congested
traffic states.
Recent theoretical studies using one lane models
\cite{Helbingcondmat,Lee99E}
suggest that the flux level $f_{\rm up}$ at the right upstream of the 
congested region,
where the free flow is maintained, and the on-ramp flux $f_{\rm rmp}$ are
the two important control parameters.
These studies assume an ideal situation where $f_{\rm up}$
and $f_{\rm rmp}$ are constants. While they
fluctuate in reality, however, 
dominant fluctuations
come from short time scale (one minute) variations and 
the fluctuations are greatly
suppressed in long time scale (ten minutes or longer).
Thus we use the average values of $f_{\rm up}$ and $f_{\rm rmp}$
over the time intervals (typically 1 hour long)
during which a particular state is maintained \cite{onrampcomment}.
In Fig.~\ref{fig:phase diagram},
each point ($f_{\rm rmp}$, $f_{\rm up}$) thus obtained
is marked with a different symbol
depending on the maintained congested traffic state. 
Notice that 
although there are some overlaps,
each symbol occupies a clearly distinguishable region in the 
$f_{\rm rmp}$-$f_{\rm up}$ plane.
This difference in the data locations verifies the roles of 
$f_{\rm rmp}$ and $f_{\rm up}$
as important control parameters, in agreement with Refs. 
\cite{Helbingcondmat,Lee99E}.
Also the metastability between the free flow and the CT1, CT2, CT3 states
is observed as studied in Ref.\cite{Lee99E}.
 
 We now make a detailed comparison of the measurement data with 
the theoretical studies \cite{Helbingcondmat,Lee99E}.
The CT1 state is similar to the recurring hump (RH) state in 
Refs. \cite{Lee98L,Lee99E}.
In both states, the congested regions do not expand
and systematic oscillations develop.
The oscillation in the CT1 state, however, exhibits features that
are not shared by the RH state, such as the oscillation amplification
and the period enhancement \cite{oscampcomment}.
The CT2 state can be related to the pinned localized cluster (PLC)
state \cite{Helbingcondmat,Lee99E} [in Ref.~\cite{Lee99E},
a different term ``standing localized cluster'' (SLC) state is used to 
denote the same state].
In both states, the congested region does not expand and 
no systematic oscillation develops.
Also the spatial variation 
of the long time ($\sim$ 1 hour) averaged density-flow relations
from the upstream to the downstream  
is essentially identical to the pattern for the PLC state
[Fig.~2(b) in Ref. \cite{Lee99E}].
And the data locations of the CT2 state in the $f_{\rm rmp}$-$f_{\rm up}$
plane [Fig.~\ref{fig:phase diagram}] are to the left of those
of the CT1 state, which
agrees with the relationship between the PLC and the RH state \cite{Lee99E}.
The two-dimensional covering of the density-flow plane 
in the CT1 and CT2 states, on
the other hand, is not shared by the RH and PLC states.
We speculate that the covering property may be due to fluctuations
effects that are not taken into account in Refs.~\cite{Helbingcondmat,Lee99E}.
For example, it is recently demonstrated that 
fluctuations in vehicle types can generate 
the two-dimensional covering \cite{HelbingJPA}. 
We also suspect that short time scale fluctuations 
in $f_{\rm up}$ and $f_{\rm rmp}$
may generate a similar effect.

 The CT3 state is also similar to the predicted PLC state
in regard to the stationary boundaries of the congested region
and the absence of the systematic oscillations.
The data locations of the state in the $f_{\rm up}$-$f_{\rm rmp}$
plane [Fig.~\ref{fig:phase diagram}] are also
in reasonable agreement with those of the PLC state 
\cite{Helbingcondmat,Lee99E}.
However the property of the constant velocity in this
state is not shared by the PLC state.
This property implies that the car following dynamics
in this state is significantly different from that
assumed in many theoretical models 
\cite{KNagel,HelbingL},
that is, the velocity adjustment to the spatial gap. 

 The CT4 state can be related to the oscillating congested traffic (OCT) 
state or homogeneous congested traffic (HCT) state
\cite{Helbingcondmat,Lee99E},
both of which exhibit the expansion of the congested region to the upstream.
In theoretical studies, it is found that the congested region of the OCT
state contains large clusters with the jam character while the congested region
of the HCT state is homogeneous \cite{Helbingcondmat,Lee99E}. 
The density-flow relation of the CT4 state does not demonstrate
the characteristic line of the jam, which suggests that 
the HCT state is the proper theoretical counterpart of the CT4 state.
As for the two-dimensional covering property of the CT4 state,
we mention that the same covering can be reproduced for the HCT state 
\cite{HelbingJPA}. 
Also the data locations of the CT4 state in the phase diagram are consistent
with the prediction for the HCT state.

 We next compare our data with the German highway
data in Ref.~\cite{Kerner96E1},
where congested traffic flows are classified into 3 types
(i,ii,iii)
according to local density-flow relations without much regard to
spatial structures.
In this classification, the CT1, CT2, and CT4 states with 
nonstationary density-flow relations belong
to the type (iii), and the CT3 state with the stationary velocity
property to the type (ii).
We also observe the type (i) state
that is characterized by the stationary velocity and density profiles.
However, this state is always short-lived (less than 5 minutes),
which is too short for our analysis,
and we do not include this state in our classification.

 It is also interesting to compare the CT1 state with the congested
traffic state reported in Ref.~\cite{Kerner98L}, which we call 
CT1' state tentatively. 
In both states, the amplification and the periodicity
enhancement of the velocity oscillation occur and  
the average velocity levels rise during the amplification.
On the other hand, the velocity oscillation in the CT1' state grows to generate
mature jam clusters and the density-flow
relation approaches the characteristic line 
(the line J in Ref.~\cite{Kerner98L})
of the traffic jam, while this feature is absent in the CT1 state
[see Fig.~\ref{fig:roadconf}(d)].
This difference suggests that the CT1 and CT1' states may be
distinct congested traffic states.
They also seem to appear at different regions in the 
$f_{\rm rmp}$-$f_{\rm up}$ plane.
From the data give in Ref.~\cite{Kerner98L}, we estimate $f_{\rm up} \geq$
1600 veh/h (information for $f_{\rm rmp}$
is not available) for the CT1' state
(compare with Fig.~\ref{fig:phase diagram}).
A recent study \cite{Helbingcondmat99} also suggests that
the CT1' state is related rather to the theoretically predicted OCT state
\cite{Helbingcondmat,Lee99E}, instead of the RH state.

 In summary, 4 congested traffic states are identified
by combining temporal traffic patterns at fixed locations
and spatial structure of the congested region.
It is found that these 4 states appear at different
levels of the upstream flux and the on-ramp flux.
An empirical phase diagram is constructed and compared with
recent predictions.
Many properties of the observed states agree with predictions
but deviations are also found.
Lastly we mention that there exist regions in the
$f_{\rm rmp}$-$f_{\rm up}$ plane which are not probed by our data.
Thus it is possible that additional congested
traffic states exist in those regions. Further investigation
is necessary.

 We thank Young-Ihn Lee and Seung Jin Lee for providing the traffic data,
and Sung Yong Park for fruitful discussions.
H.-W.L. was supported by the Korea Science and Engineering Foundation.
This work is supported by the Korea Science and Engineering
Foundation through the SRC program at SNU-CTP,
and also by Korea Research Foundation (1998-015-D00055).

\end{multicols}


\begin{thebibliography}{7}
\bibitem{CTcomment} We use the term congested traffic states to denote
all states which do not belong to the free flow.
\bibitem{dividercomment} To be precise, the lane divider
is not perfect at two locations where a small number of vehicles
can be exchanged between the lanes 2 and 3.
\bibitem{onrampcomment}Both quantities are averaged over the two inner lanes.
And $f_{\rm rmp}$ is estimated from the difference between
the maximum flux level near the downstream end of the congested region
and the flux level near the upstream end of congested region.
\bibitem{oscampcomment} A recent theoretical study \cite{Helbingcondmat99}
reproduces the oscillation amplification and the
period enhancement.
But in this study, the congested region expands monotonically with time,
which makes a clear distinction from the CT1 state.
\end{thebibliography}
\end{document}